\documentclass[fleqn,usenatbib]{mnras}

\usepackage[T1]{fontenc}

\DeclareRobustCommand{\VAN}[3]{#2}
\let\VANthebibliography\thebibliography
\def\thebibliography{\DeclareRobustCommand{\VAN}[3]{##3}\VANthebibliography}

\usepackage{graphicx}	
\usepackage{amsmath}	

\newcommand{\mjup}{\,$M_{\rm J}$}
\newcommand{\rjup}{\,$R_{\rm J}$}
\newcommand{\kepler}{{\it Kepler}}

\newcommand{\tess}{{\it TESS}}

\newcommand{\gaia}{{\it Gaia}}

\newcommand{\ngts}{{NGTS}}

\newcommand{\coralie}{{CORALIE}}

\newcommand{\teff}{{T$_{\rm eff}$}}
\newcommand{\logg}{{$\log$ g}}
\newcommand{\feh}{[Fe/H]}
\newcommand{\vsini}{$V\sin i$}
\newcommand{\system}{{\rm TIC-320687387 AB}}
\newcommand{\systemt}{{\rm TIC-320687387}}
\newcommand{\systemA}{{\rm TIC-320687387 A}}
\newcommand{\systemB}{{\rm TIC-320687387 B}}

\usepackage{newtxtext,newtxmath}

\title[\system]{\systemB: a long-period eclipsing M-dwarf close to the hydrogen burning limit}
\author[Gill et al.]{
Samuel Gill,$^{1,2}$
Sol\`ene Ulmer-Moll,$^{3}$
Peter~J.~Wheatley,$^{1,2}$
Daniel Bayliss,$^{1,2}$
Matthew~R.~Burleigh,$^{4}$
\newauthor Jack~S.~Acton,$^{4}$
Sarah~L.~Casewell,$^{4}$
Christopher A. Watson,$^{11}$
Monika Lendl,$^{3}$ Hannah~L.~Worters,$^{5}$
\newauthor Ramotholo~R.~Sefako,$^{5}$
David~R.~Anderson,$^{1,2}$
Douglas R. Alves,$^{6}$
Fran\c{c}ois Bouchy,$^{3}$
Edward~M.~Bryant,$^{1,2}$
\newauthor Philipp Eigm\"uller,$^{12}$
Edward Gillen,$^{8,9}$\thanks{Winton Fellow},
Michael~R.~Goad,$^{4}$
Nolan Grieves,$^{3}$
Maximilian~N.~G{\"u}nther,$^{15}$\thanks{ESA Research Fellow},
\newauthor Beth~A.~Henderson,$^{4}$
James S. Jenkins,$^{6,7}$
Lokesh Mishra,$^{3}$
Maximiliano Moyano,$^{10}$
Hugh~P.~Osborn,$^{13,14}$
\newauthor Rosanna~H.~Tilbrook,$^{4}$
St\'ephane Udry,$^{3}$
Jose I. Vines,$^{6}$
Richard~G.~West$^{1,2}$\\
$^{1}$ Department of Physics, University of Warwick, Gibbet Hill Road, Coventry CV4 7AL, UK \\
$^{2}$ Centre for Exoplanets and Habitability, University of Warwick, Gibbet Hill Road, Coventry CV4 7AL, UK \\ 
$^{3}$Observatoire de Gen{\`e}ve, Universit{\'e} de Gen{\`e}ve, Chemin Pegasi 51, 1290 Sauverny, Switzerland \\
$^{4}$School of Physics and Astronomy, University of Leicester, Leicester, LE1 7RH, UK\\
$^{5}$South African Astronomical Observatory, P.O Box 9, Observatory 7935, Cape Town, South Africa\\
$^{6}$Departamento de Astronom\'ia, Universidad de Chile, Casilla 36-D, Santiago, Chile\\
$^{7}$ N\'ucleo de Astronom\'ia, Facultad de Ingenier\'ia y Ciencias, Universidad Diego Portales, Av. Ej\'ercito 441, Santiago, Chile\\
$^{8}$Astronomy Unit, Queen Mary University of London, Mile End Road, London E1 4NS, UK\\
$^{9}$Astrophysics Group, Cavendish Laboratory, J.J. Thomson Avenue, Cambridge CB3 0HE, UK\\
$^{10}$ Instituto de Astronom\'ia, Universidad Cat\'olica del Norte,Angamos 0610, 1270709, Antofagasta, Chile\\
$^{11}$ Astrophysics Research Centre, School of Mathematics and Physics, Queen’s University Belfast, BT7 1NN, Belfast, UK \\
$^{12}$Institute of Planetary Research, German Aerospace Center, Rutherfordstrasse 2., 12489 Berlin, Germany\\
$^{13}$ NCCR/PlanetS, Centre for Space \& Habitability, University of Bern, Bern, Switzerland \\
$^{14}$ Department of Physics and Kavli Institute for Astrophysics and Space Research, Massachusetts Institute of Technology, Cambridge, MA 02139, USA \\
$^{15}$ European Space Agency (ESA), European Space Research and Technology Centre (ESTEC), Keplerlaan 1, 2201 AZ Noordwijk, The Netherlands \\
}

\date{Accepted XXX. Received YYY; in original form ZZZ}

\pubyear{2015}

\begin{document}
\label{firstpage}
\pagerange{\pageref{firstpage}--\pageref{lastpage}}
\maketitle

\begin{abstract}
We are using precise radial velocities from \coralie\ together with precision photometry from the Next Generation Transit Survey (\ngts) to follow up stars with single-transit events detected with the Transiting Exoplanet Survey Satellite (\tess).
As part of this survey we identified a single transit on the star \systemt, a bright ($T=11.6$) G-dwarf observed by \tess\ in Sector 13 and 27.  From subsequent monitoring of \systemt\ with \coralie, \ngts, and Lesedi we determined that the companion, \systemB, 
is a very low-mass star with a mass of $96.2 \pm _{2.0}^{1.9}$ \mjup\ and radius of $1.14 \pm _{0.02}^{0.02}$ \rjup\ placing it close to the hydrogen burning limit ($\sim 80$\,\mjup). 
\systemB\ has a wide and eccentric orbit, with a period of 29.77381 days and an eccentricity of $0.366 \pm 0.003$. 
Eclipsing systems such as \system\ allow us to test stellar evolution models for low-mass stars, which in turn are needed to calculate accurate masses and radii for exoplanets orbiting single low-mass stars. 
The wide orbit of \systemB\ makes it particularly valuable as its evolution can be assumed to be free from perturbations caused by tidal interactions with its G-type host star.

\end{abstract}




\section{Introduction}
The small size and low luminosity of late M-dwarfs make them ideal targets to detect temperate terrestrial planets, such as those found in the TRAPPIST-1 system \citep{2016Natur.533..221G,2017Natur.542..456G}. Planet occurrence rates for M-dwarf hosts from \kepler\  also appear to be higher than for FGK hosts \citep[e.g.][]{2020MNRAS.498.2249H}, and finding transiting planets around M-dwarfs is a key goal of the \tess\ mission \citep{2015JATIS...1a4003R} as well as ground-based surveys such as MEarth \citep{2008PASP..120..317N} and SPECULOOS \citep{2021A&A...645A.100S}. \tess\ has already found 49 planets with M-dwarf host stars.\footnote{exoplanetarchive.ipac.caltech.edu - 2021-12-10}

As with all transiting exoplanets, however, the planetary mass and radius can only be measured with respect to the properties of the host star. It is a concern, therefore, that observations of eclipsing M-dwarfs often reveal them to be cooler and larger than predicted by stellar models \citep[e.g.][and references therin]{2017ApJ...844..134L,2018MNRAS.481.1083P}. In order to accurately characterise the population of temperate exoplanets, it is therefore imperative that we understand these low mass stars as completely as possible. 

The tension between measured M-dwarf physical properties and models appears to span the spectral type with no obvious deviations around the transition from partially to fully convective stars \citep{2018MNRAS.481.1083P}. Modification of stellar convection by magnetic fields is often invoked to explain these discrepancies \citep[e.g.][]{2007AA...472L..17C,2013EAS....64..127F}, with large starspot fractions leading to cooler measured temperatures and radius inflation compensating for the lower flux from the photosphere. There have also been suggestions that the level of radius inflation is related to metallicity \citep[e.g.][]{2000ApJ...535..965L,2006ApJ...644..475B,2007ApJ...660..732L}.

Much of the evidence for oversized M-dwarfs comes from eclipsing binary systems, where masses and radii can be measured precisely. These low-mass eclipsing binaries (EBLMs) have been found in large numbers with ground-based transit surveys such as WASP \citep{2006PASP..118.1407P,2017A&A...608A.129T,2019A&A...626A.119G}. However, due to transit geometry, the known population of EBLMs is strongly weighted to short-period close binaries, where strong tidal interactions can maintain rapid rotation leading to enhanced magnetic activity \citep[e.g.][]{2011ApJ...728...48K}. This makes it difficult to separate single-star evolution from tidal effects experienced only in close binaries. 

One way to find longer-period wider-separation EBLMs that are free of tidal interactions is to exploit single-transit events detected with the \tess\ mission. We have begun a programme of photometric and spectroscopic follow-up of \tess\ single transit events, finding a mixture of long-period exoplanets \citep[e.g.][]{2020ApJ...898L..11G} and low-mass stellar companions \citep[e.g.][]{2020MNRAS.491.1548G,2020MNRAS.495.2713G,2020MNRAS.492.1761L,2021A&A...652A.127G}.

In this paper, we present the orbital solution of the \tess\ single-transit candidate \system, which we have found to be a G2+M7 binary on a 30-day orbit. In Sect.\,\ref{sec:observations}, we describe the identification of the single-transit event as part of our warm Jupiter program, and we detail the observations required to measure the orbital solution. In Sect.\,\ref{sec:analysis} we describe our modelling of the \system\ system, while in Sect.\,\ref{sec:discussion} we discuss our results and the implications for radius inflation in late-type M-dwarfs.

\section{Observations}\label{sec:observations}
\subsection{\tess\ single-transit detection}\label{sec:single_transit}

\begin{table}
\caption{Photometric colours, stellar atmospheric parameters, and physical properties of \systemA.}              
\label{tab:systemAparameters}      
\centering   
\begin{tabular}{l c c}          
\hline\hline                        
Parameter & value & Source\\
\hline 
Gaia eDR3 Source ID & 6641131183310690432 & 1  \\
RA  &  $19^{\rm h} 51^{'}18.41^{"}$ & 1  \\
Dec  & $-55^{\circ} 32' 47^{"}$ & 1  \\

\\
pmRA [$\rm mas\, \rm yr^{-1}$] & $31.827 \pm 0.060$  & 1  \\
pmDec [$\rm mas\, \rm yr^{-1}$] & $-35.648 \pm 0.048$  & 1  \\
Parallax [$\rm mas$] & $2.9720 \pm  0.0351$  & 1  \\ 
Distance [pc] & $336.5 \pm 4.0$  & 1  \\ \\
Magnitudes\\
GAIA G & $12.0148 \pm 0.0002$  & 1  \\
GAIA BP & $12.3383 \pm 0.0012$  & 1  \\
GAIA RP & $11.5246 \pm 0.0009$  & 1  \\
TESS [T]  & $11.591 \pm 0.006$ & 2\\
APASS9 [B] & $13.012 \pm 0.398$ & 3\\
APASS9 [V] & $12.238 \pm 0.028$ & 3\\
2MASS [J]  & $10.955 \pm 0.022$ & 4 \\
2MASS [H]  & $10.681 \pm 0.026$ & 4 \\
2MASS [K$_{s}$]  & $10.624 \pm 0.019$ & 4 \\ \\

Spectroscopic parameters \\
$\rm T_{\rm eff}$ $\rm(K)$ & $5780 \pm 80$ & 5 \\
$\log g$ (dex)  & $4.4 \pm 0.1$ & 5 \\
$\xi_{\rm t}\, (\rm km\,s^{-1})$ & $1.15 \pm 0.18^6$ & 5 \\
$v_{\rm mac}\, (\rm km\,s^{-1})$ & $3.97 \pm 0.73^6$ & 5 \\
Vsin$i$ (km\,s$^{-1}$) & $2.5 \pm 0.8$ & 5 \\
$\rm [Fe/H]$ (dex) & $0.30 \pm 0.08$ & 5 \\ \\

Host parameters \\
$M_{\rm A}$ [$M_{\odot}$] & $1.080 \pm 0.034$ & 5 \\
$R_{\rm A}$ [$R_{\odot}$] & $1.158 \pm 0.016$ & 5 \\
Age [Gyr] & $5 \pm 3$ & 5 \\
\hline
\multicolumn{3}{l}{$^1$ \citet{2018A&A...616A...1G}, $^2$ \citet{2018AJ....156..102S},}\\
\multicolumn{3}{l}{$^3$ \citet{2015AAS...22533616H},  $^4$ \citet{2006AJ....131.1163S},}\\
\multicolumn{3}{l}{$^5$ this work, $^6$ uncertainties from \citet{2015PhDT........16D}.}
\end{tabular}
\end{table}

\begin{figure}
    \centering
    \includegraphics[width=0.45\textwidth]{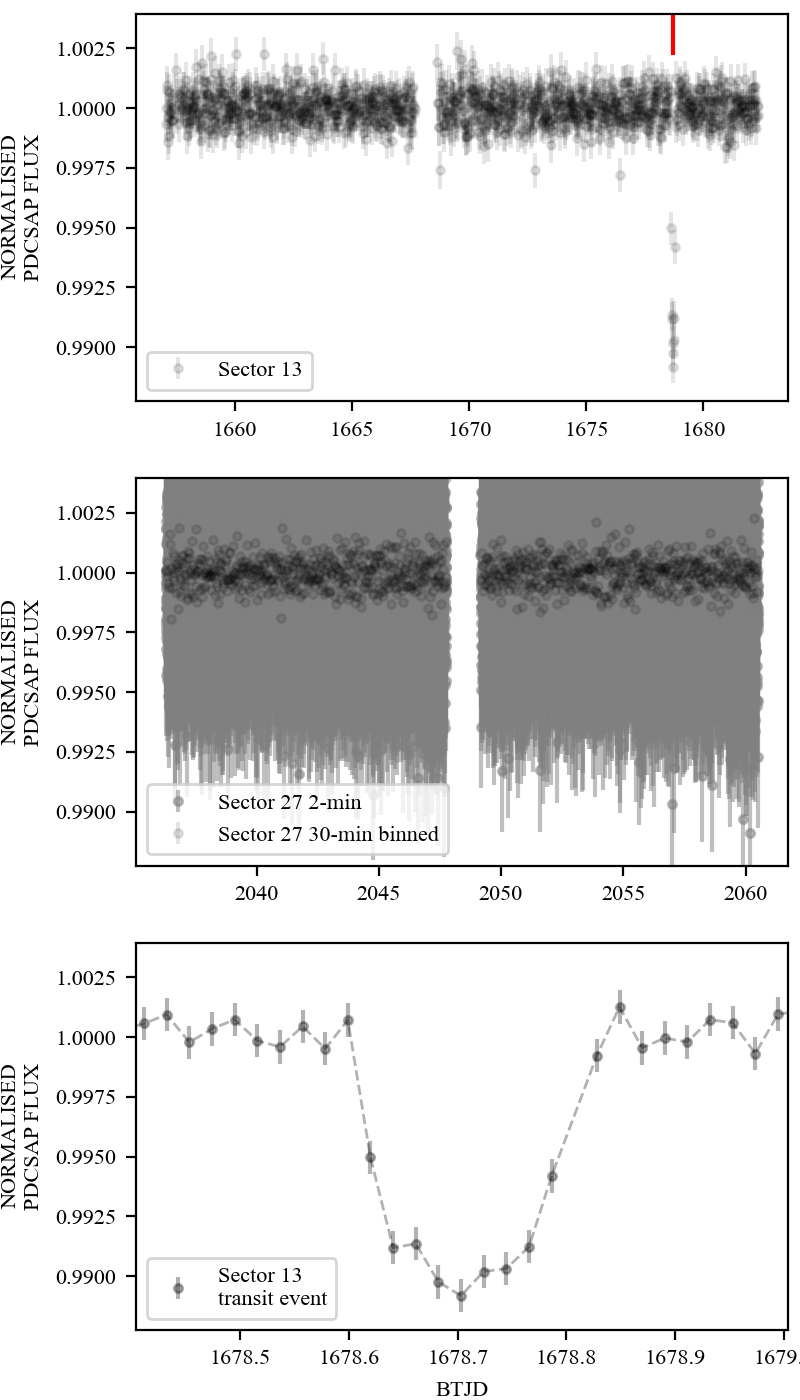}
    \caption{\tess\ SPOC lightcurves for sector 13 (30-minute cadence; upper panel) with the single transit event marked (red) along with sector 27 (2-minute cadence; middle panel). A closer look at the single-transit event in sector 13 is shown in the lower panel.}
    \label{fig:figure_1}
\end{figure}

We searched our own \tess\ full-frame lightcurves for single-transit events as described in \citep{2020MNRAS.491.1548G,2020MNRAS.495.2713G,2020ApJ...898L..11G}. \systemt\  was observed with Camera 2 during Sectors 13 (2019-Jun-19 to 2019-Jul-18) and 27 (2020-Jul-04 to 2020-Jul-30). We identified a single transit event 
in our search of \tess\ Sector 13 data at JD\,2458678.70408.  \systemA\ is a $T=11.6$ G-dwarf, and we list its stellar parameters in Table \ref{tab:systemAparameters}.  The transit depth was 11\,ppt and is clearly significant compared with the out-of-transit lightcurve scatter of r.m.s = 0.1\,ppt (see Figure~\ref{fig:figure_1}). We carefully inspected calibrated TESS full-frame images for signs of asteroids, spacecraft jitter, stray light variations, or background eclipsing binaries.  We found no evidence to suggest the event was anything other than a real astrophysical single-transit.  

\tess\ photometry of \system\ was also processed by the Science Processing Operations Center \citep[SPOC; ][]{2016SPIE.9913E..3EJ} and made publicly available on the Mikulski Archive for Space Telescopes (MAST)\footnote{https://mast.stsci.edu/}. Measurements of Sector 13 was made at 30-minute cadence and sector 27 at 2-minute cadence. We downloaded the \tess\ SPOC HLSP data \citep{2020RNAAS...4..201C} from MAST which included Simple Aperture Photometry (SAP) extracted from the pipeline-derived photometric aperture \citep{2010SPIE.7740E..23T,2017ksci.rept....6M} along with the Presearch Data Conditioning SAP (PDCSAP) light curve, which has been corrected for systematic trends shared by other stars on the detector (co-trending basis vectors). This product is significantly cleaner than its SAP counterpart and corrected for dilution so we present the analysis of the PDCSAP lightcurve in this work (Figure \ref{fig:figure_1}).


\subsection{Spectroscopic follow up with \coralie}

\begin{table}
\caption{Radial velocity observations of \system\ and their associated errors from \coralie.}              
\label{tab:radial_velocities}      
\centering   
\begin{tabular}{l c}          
\hline\hline                        
JD & Radial velocity [$\rm km\, \rm s^{-1}$]\\
\hline 
2459323.873253 & $-50.0899 \pm 0.0391$ \\
2459332.875484 & $-49.3755 \pm 0.0486$ \\
2459345.853170 & $-55.0933 \pm 0.0519$ \\
2459350.850859 & $-52.0922 \pm 0.0851$ \\
2459363.842946 & $-52.7297 \pm 0.0343$ \\
2459368.741216 & $-57.7449 \pm 0.0300$ \\
2459375.766215 & $-54.9970 \pm 0.0323$ \\
2459415.760069 & $-48.0367 \pm 0.0448$ \\

\hline
\end{tabular}
\end{table}


We made high-precision radial-velocity measurements of \system\ using \coralie\ --- a fiber-fed \'{e}chelle spectrograph installed on the 1.2-m Leonard Euler telescope at the ESO La Silla Observatory \citep{2001A&A...379..279Q}.
A total of 8 spectra were obtained between 2021 April 19 and 2021 July 20, each with an exposure time of 2400\,s. The spectra were reduced using the standard \coralie\ reduction pipeline, and radial velocity measurements derived from cross-correlation with a numerical G2 mask. These observations are summarised in Table~\ref{tab:radial_velocities} and plotted in Figure~\ref{fig:fig4} showing the best-fitting orbital solution. These data have a high radial velocity semi-amplitude  consistent with a low-mass stellar companion on an eccentric orbit. We inspected potential dependencies between radial velocities and bisector spans and found little evidence of correlation. The radial velocity variations measured by \coralie\ and the initial \tess\ transit were sufficient to determine the approximate orbital period of \system.

\subsection{Transit photometry with \ngts}

\begin{figure}
    \centering
    \includegraphics[width=0.45\textwidth]{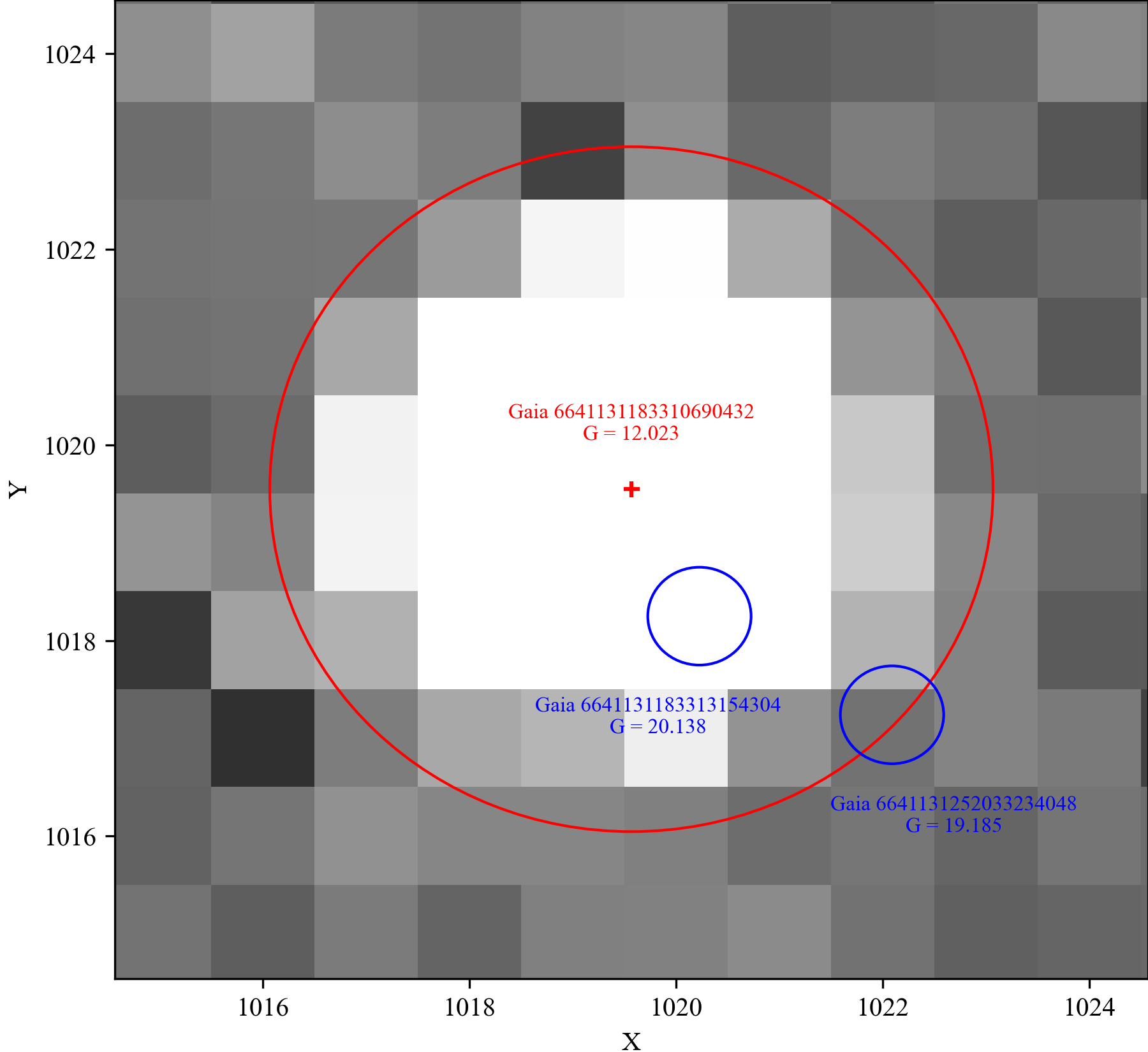}
    \caption{The \protect \ngts\ reference image with the 3.5-pixel aperture plotted (red line) around \protect \system\ (red cross) along with nearby stars from Gaia eDR3 (blue circles) with corresponding \gaia\ IDs and \gaia\ G magnitudes.}
    \label{fig:fig2}
\end{figure}

We also carried out photometric monitoring of  \system\ from the night of 2021 May 8 using a single telescope of the Next generation Transit Survey \cite[NGTS; ][]{2018MNRAS.475.4476W}, which is located at the ESO Paranal Observatory in Chile.   
Each \ngts\ telescope has a field-of-view of 8 square degrees, providing sufficient reference stars for even the brightest TESS candidates. The telescopes have apertures of 20\,cm and observe with a custom filter between 520-890\,nm. We observed \system\ with 10-s exposures when the airmass was below 2.5.  Data were reduced on-site the following day using standard aperture photometry routines. We used the template matching algorithm described in \citet{2020MNRAS.491.1548G} to automatically search newly obtained \ngts\ photometric observations for transit events. In total, 121,738 photometric measurements of \system\ were made over 121 nights.

We detected two transit events on \system\ with \ngts. The first was an ingress event on the night of 2021 August 26 with a significance of $\Delta \log \mathcal{L} = 1244$, shown in the second panel of Figure \ref{fig:fig4}. The second transit event was on 2021 September 24 with $\Delta \log \mathcal{L} = 3355$, shown in the third panel of Figure \ref{fig:fig4}. This event constrains the transit width and impact parameter resulting in a more precise stellar density that otherwise possible with \tess\ alone. There are 2 faint sources that reside within the Gaia aperture (see Figure \ref{fig:fig2}). Both sources together contribute 0.192\% of the total flux in the aperture and cannot be the source of the transit events.

\subsection{Transit photometry with Lesedi}
On 2021 October 24 we also managed to observe a fourth transit event using Lesedi, a new 1-m telescope, at the South African Astronomical Observatory (SAAO). We obtained 1145 consecutive $8$~second $V$~band images with the Sutherland High-speed Optical Camera, SHOC \citep{2013PASP..125..976C}, a frame-transfer CCD camera with a $5.72 \times 5.72$~arcminutes field of view (plate scale of $0.335$~arcsec/pix), for a total observation time of $9160$~seconds ($2.54$~hours). Conditions during the observation were clear, with seeing improving from $2$ to $1.5$~arcseconds, and $\approx50$\% humidity. The images were bias and flat-field corrected using the local \textsc{python}-based SHOC pipeline, which utilises \textsc{iraf} photometry tasks (\textsc{pyraf}). We performed aperture photometry on the target and two comparison stars using the Starlink package \textsc{autophotom}. A $4$~pixel radius aperture was selected to maximise the signal-to-noise. The comparison stars were combined to perform differential photometry on the target.  The resulting light curve is presented in Figure~\ref{fig:fig4}.

\section{Analysis}\label{sec:analysis}

\subsection{Stellar atmospheric and physical parameters}

\begin{figure*}
    \centering
    \includegraphics[width=0.9\textwidth]{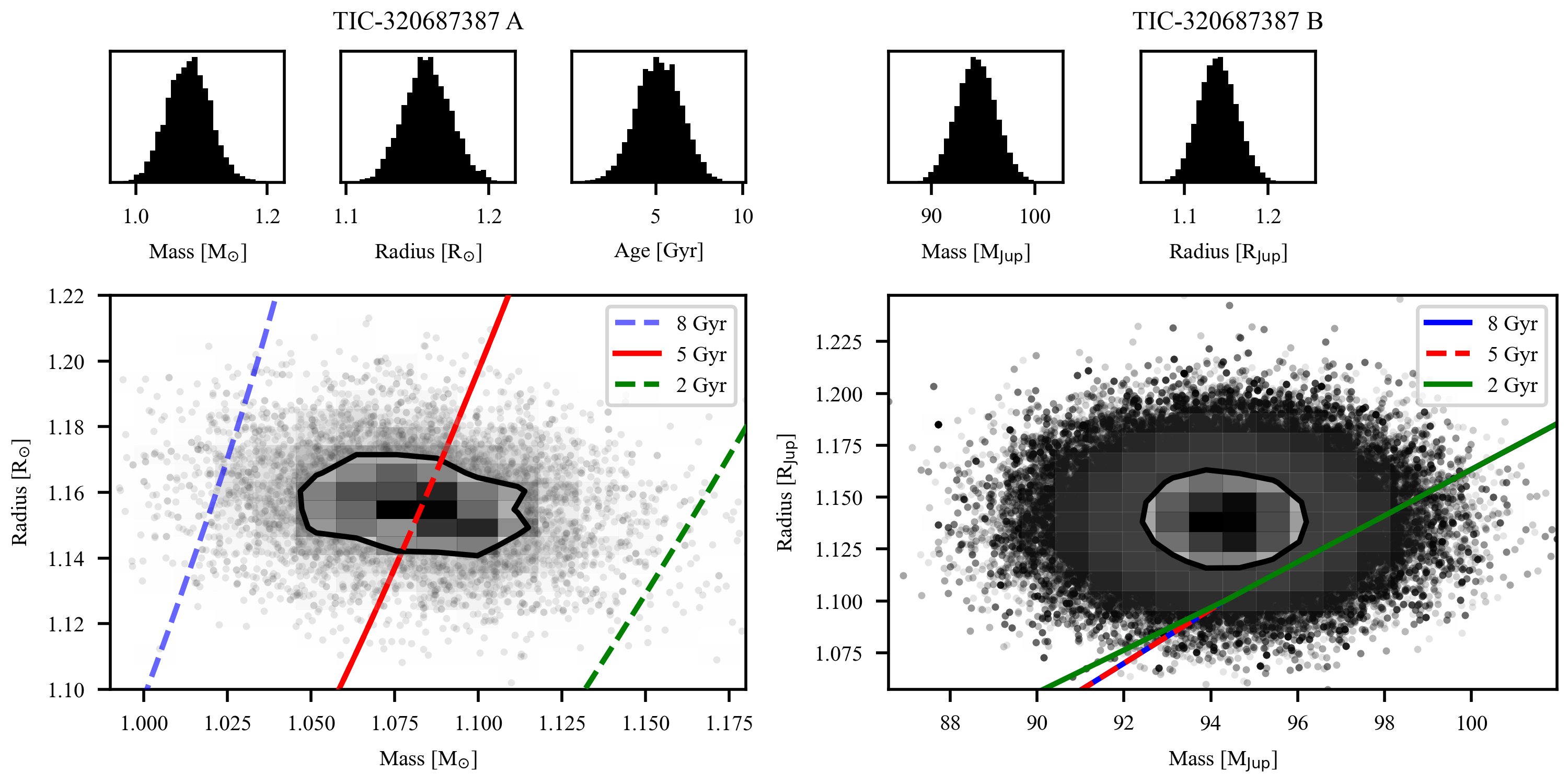}
    \caption{Upper panels show posterior probability distributions for mass, radius, and age of \systemA\ (left panels) and \systemB\ (right panels). Lower panels show 2d mass-radius posterior probability distributions with a 1-$\sigma$ contour (black line) along with isochrones for 2\,Gyr, 5\,Gyr, and 8\,Gyr from MESA (\systemA) and BHAC15 (\systemB).} 
    \label{fig:primary_star}
\end{figure*}

We corrected each \coralie\ spectrum into the laboratory reference frame using radial velocities from Table \ref{tab:radial_velocities} and co-added them onto a common wavelength scale to create a high quality spectrum with S/N $\sim 35$. As described by \citet{2020ApJ...898L..11G}, a grid of pre-computed model spectra were synthesised with the software package \textsc{spectrum} \citep{1999ascl.soft10002G} using MARCS model atmospheres \citep{2008A&A...486..951G}, version 5 of the GAIA ESO survey (GES) atomic line list and solar abundances from \citet{2009ARA&A..47..481A}. Values of macroturbulence ($v_{\rm mac}$) and microturbulence ($\xi_{\rm t}$) were calculated using equations 5.10 and 3.1 respectively from \citet{2015PhDT........16D}. Given these models, we used the H$\alpha$, NaI\,D, and MgI\,b lines to determine the stellar effective temperature, \teff, and surface gravity, \logg. Individual FeI and FeII lines provided a measurement of metallicity, \feh, and the rotational broadening projected into the line of sight, \vsini. 

We used the method described in \citet{2020MNRAS.491.1548G} to determine the mass, radius, and age of \systemA.  This method uses Gaia magnitudes and parallax \citep{2018A&A...616A...1G} along with \teff\ and \feh\ from the spectroscopic analysis to determine the best-fitting stellar parameters with respect to MESA models \citep{2016ApJS..222....8D,2016ApJ...823..102C}. We found \systemA\ to be a main-sequence G-type star. Our results are in good agreement with physical parameters predicted in version 8 of the \tess\ input catalogue and the results of our analysis are shown in Figure \ref{fig:primary_star} and presented in Table \ref{tab:systemAparameters}.

\subsection{Orbital geometry and transit properties}

We modelled all photometric and radial velocity datasets simultaneously. As part of the SPOC pipeline, the PDCSAP lightcurve has been corrected for assuming a contamination ratio of 3.668\% \citep[calculated from version 8 of the \tess\ input catalogue;][]{2018AJ....156..102S}. From initial modeling we found this to be an under-correction and so we fit an additional dilution term, $l_{3,\tess}$, to the PDCSAP lightcurve. This is in line with other works \citep[e.g. ][]{2020AJ....160..153B} and suggests additional light from other sources contaminates the SPOC aperture. We found the 2 background stars in the \ngts\ aperture contribute a negligible amount of third light and do not correct the \ngts\ transit depths. 

We used the binary star model described by \citet{2020MNRAS.491.1548G} to calculate models of radial velocity and transit photometry. This utilises the analytical transit model for the power-2 limb-darkening law presented by \citet{2019A&A...622A..33M}. We fit decorrelated limb-darkening parameters $h_1$ \& $h_2$ (from Eqn. 1 \& 2 of \citealt{2018A&A...616A..39M}) with Gaussian priors centred on values interpolated from Table\,2 of \citet{2018A&A...616A..39M} using stellar atmospheric parameters from Table \ref{tab:systemAparameters} and widths of 0.003 and 0.046 respectively. The subtle differences between \tess, \ngts, and Lesedi's V-band transmission filters are such that we fitted independent values of $h_1$ and $h_2$ for each photometric dataset. The orbital period and eccentricity yield a light travel time on the order of 1-2 minutes which is significant for the cadence of \ngts\ and Lesedi observations; our model accounts for light travel time delays which causes the transits to appear early. Preliminary modelling yielded consistent transit depths across different colours and so we decide to fit a common value of $k=R_{\rm B}/R_{\rm A}$. The luminosity ratio between the host and transiting companion are such that we do not expect to see a secondary eclipse or significant dilution of the primary eclipse in our datasets (see Section \ref{discuss:secondary}).

Our model vector included the transit epoch ($T_0$), the orbital period ($P$), the scaled orbital separation ($R_{\rm A} / a$), the ratio of radii ($k = R_{\rm B} / R_{\rm A}$), the impact parameter ($b$), $l_{3,\tess}$, independent values of the photometric zero-point ($zp$), $h_1$ and $h_2$ for each photometric dataset, the radial-velocity semi-amplitude ($K_{\rm A}$), and the systematic radial velocity of the primary star ($V_0$). 
We avoid fitting the strongly correlated eccentricity ($e$) and the argument of the periastron ($\omega$) and instead used $f_c = \sqrt{e} \cos \omega$ and  $f_s = \sqrt{e} \sin \omega$ since these are less correlated and have more uniform prior probability distributions. \coralie\ radial velocity errors are occasionally underestimated in-part due to spot activity, pulsations, and granulation which can introduce noise in to the radial velocity measurements \citep{2006ApJ...642..505F}. To mitigate this, we include a jitter term, $J$, which is added in quadrature with \coralie\ radial velocity errors. 
We fit a similar term for each photometric data set, $\sigma$, which was also added in quadrature to photometric uncertainties.

The Bayesian sampler \textsc{emcee} \citep{2013PASP..125..306F} was used to explore parameter space and determine the best-fitting model for the \system\ system. We drew 100,000 steps from 46 Markov chains and discarded the first 50,000 steps as part of the burn-in phase. After visually confirming each chain had converged, we selected the trial step with the highest log-likelihood was chosen as our measurement for each fitted parameter. Asymmetric uncertainties were calculated from the difference between each measured parameter and the $16^{\rm th}$ and $84^{\rm th}$ percentiles of their cumulative posterior probability distributions.

For each valid trial step we calculate the transit width using Eqn. 3 from \citet{2003ApJ...585.1038S}. We draw random values of $M_{\rm A}$ and $R_{\rm A}$ from a normal distribution centred on measured values from Table \ref{tab:systemAparameters} with width equal to their respective uncertainties. These were combined with trial values of $P$, $e$, and $K_{\rm A}$ to make a closed-form solution of the cubic polynomial required to solve the mass function, 
\begin{equation}\label{mass_function}
\frac{(M_{\rm B} \sin i)^3}{(M_{\rm A} + M_{\rm B})^2} =  (1-e^2)^{\frac{3}{2}} \frac{P K_{\rm A}^3}{2 \pi G},
\end{equation}
for $M_{\rm B}$. The mass ratio, $q = M_{\rm B} / M_{\rm A}$, can then be used with $R_{\rm A} / a$, $f_s$, and $f_c$ to estimate the stellar density using Eqns. 1 \& 2 from \citet{2015ApJ...808..126V} along with surface gravity of the transiting companion using Eqn. 4 from \citet{2007MNRAS.379L..11S}. Trial values of $R_{\rm A}$ and $k$ were combined to calculate $R_{\rm B}$. The measured values from our joint analysis are summarised in Table \ref{tab:fitted_parameters} and Figure \ref{fig:fig4} along with derived parameters shown in Table \ref{tab:derived_parameters}.

\begin{table}
\caption{Orbital solution of the \system\ system. Asymmetric errors are reported in brackets and correspond to the difference between the median and the 16$^{th}$ (lower value) and 84$^{th}$ (upper value) percentile.}              
\label{tab:fitted_parameters}      
\centering   
\begin{tabular}{l c}          
\hline\hline                        
Parameter & value\\
\hline 
Fitted parameters \\
$\rm T_{\rm 0}$ [JD] &  $2459452.82405_{(98)}^{(98)}$\\
Period [d] & $29.77381_{(12)}^{(12)}$ \\
$R_{\rm A} / a$ & $0.0289_{(15)}^{(13)}$\\
$R_{\rm B} / R_{\rm A}$ & $0.1012_{(14)}^{(18)}$ \\
$b$ & $0.639_{(144)}^{(176)}$ \\
$\rm h_{\rm 1, \tess}$ & $0.7873_{(22)}^{(23)}$ \\
$\rm h_{\rm 2, \tess}$ & $0.4405_{(350)}^{(357)}$ \\
$\rm h_{\rm 1, \ngts}$ & $0.7638_{(23)}^{(23)}$  \\
$\rm h_{\rm 2, \ngts}$ & $0.4580_{(350)}^{(360)}$\\
$\rm h_{\rm 1, \rm Lesedi}$ & $0.7044_{(24)}^{(24)}$  \\
$\rm h_{\rm 2, \rm Lesedi}$ & $0.4681_{(364)}^{(365)}$\\
$zp_{\tess}$ & $1.00013_{(19)}^{(20)}$ \\
$zp_{\ngts}$ & $0.99990_{(13)}^{(12)}$ \\
$zp_{\rm Lesedi}$ & $0.99976_{(14)}^{(14)}$ \\
$l_{3,\tess}$ & $0.110_{(50)}^{(52)}$ \\
$\sigma_{\rm TESS}$  & $0.00014_{(13)}^{(9)}$ \\
$\sigma_{\rm NGTS}$  & $0.00047_{(21)}^{(24)}$ \\
$\sigma_{\rm Lesedi}$  & $0.00441_{(8)}^{(8)}$ \\
$K_{\rm A}$ [km\,s$^{-1}$] & $5.983_{(26)}^{(27)}$ \\
$f_s$ & $0.563_{(3)}^{(3)}$ \\
$f_c$ & $0.220_{(4)}^{(4)}$ \\
$V_0$ [km\,s$^{-1}$] & $-52.576_{(12)}^{(13)}$ \\
$J$ [km\,s$^{-1}$] & $0.016_{(8)}^{(11)}$ \\
\hline
\end{tabular}
\end{table}

\subsection{Star spot modulation}

\begin{figure}
    \centering
    \includegraphics[width=0.45\textwidth]{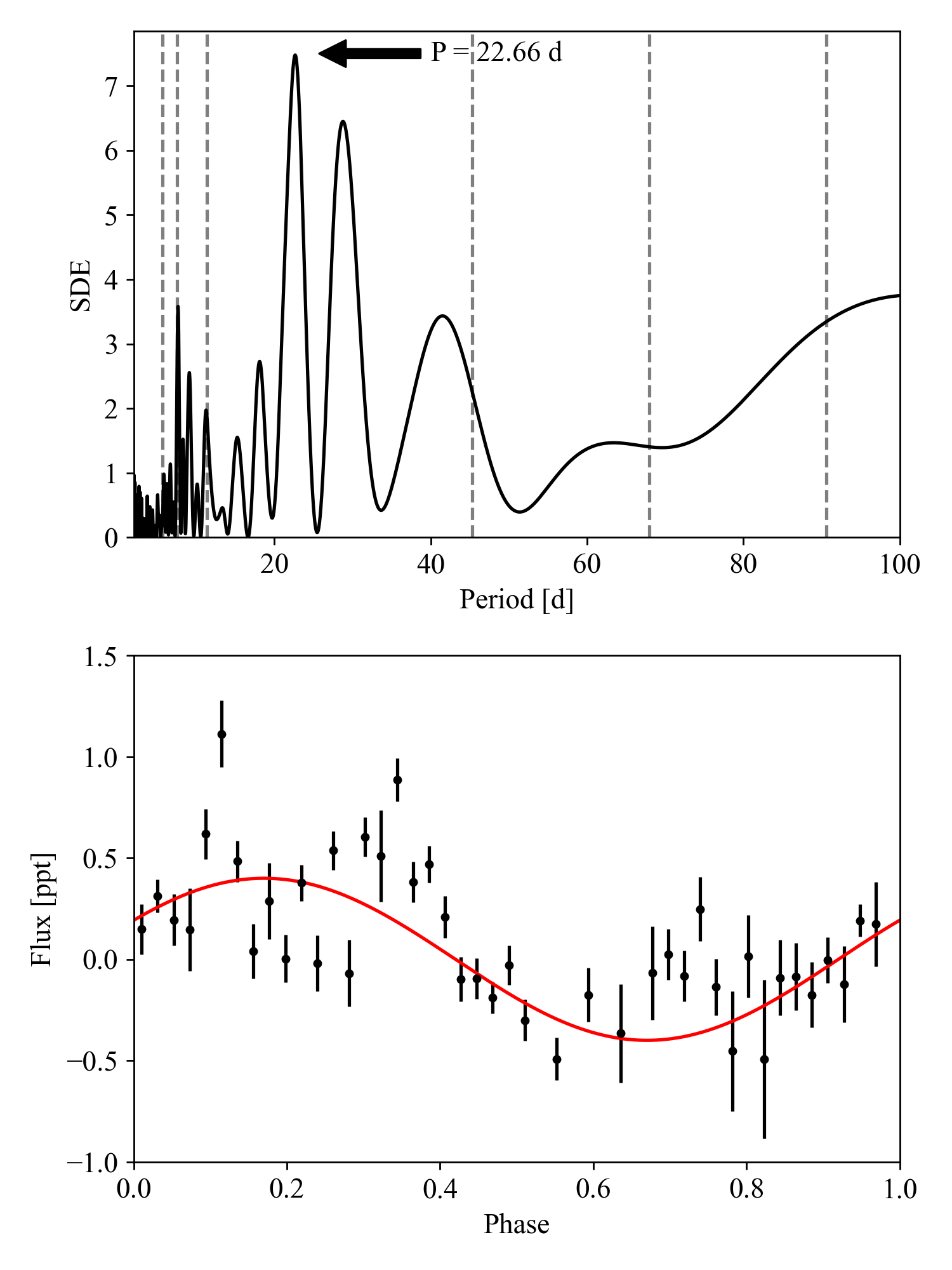}
    \caption{Upper panel - Lomb-scargle diagram for the out of transit \ngts\ photometery (black) with peaks marking harmonics (grey vertical lines). The peak of the Lomb-scargle periodogram (22.53-day period) is marked with a black arrow. Lower panel - phased and binned \ngts\ photometry on the 22.53-day period with best fitting sinusoid (red) indicating a 0.4\,ppt modulation likely caused by star spots and stellar rotation. }
    \label{fig:LS}
\end{figure}

Independent measurements of the rotation period came from \ngts\ photometry which show a subtle brightening/dimming effect as star spots come in and go from the facing hemisphere of \systemA. The longevity of spots is a limiting factor for measuring the rotational period and we assume that the average spot lifetime exceeds the rotation period of \systemA. 
A Lomb-Scargle analysis of the out-of-transit \ngts\ photometry for \systemt\ reveals a significant ($\rm sde > 8$) peak at 22.53 days  (Figure \ref{fig:LS}) corresponding to 0.4\,ppt variation equivalent to a rigid body rotational velocity of $\sim 2.6\, \rm km\, \rm s^{-1}$.

\section{Discussion}\label{sec:discussion}

\begin{figure}
    \centering
    \includegraphics[width=0.47\textwidth]{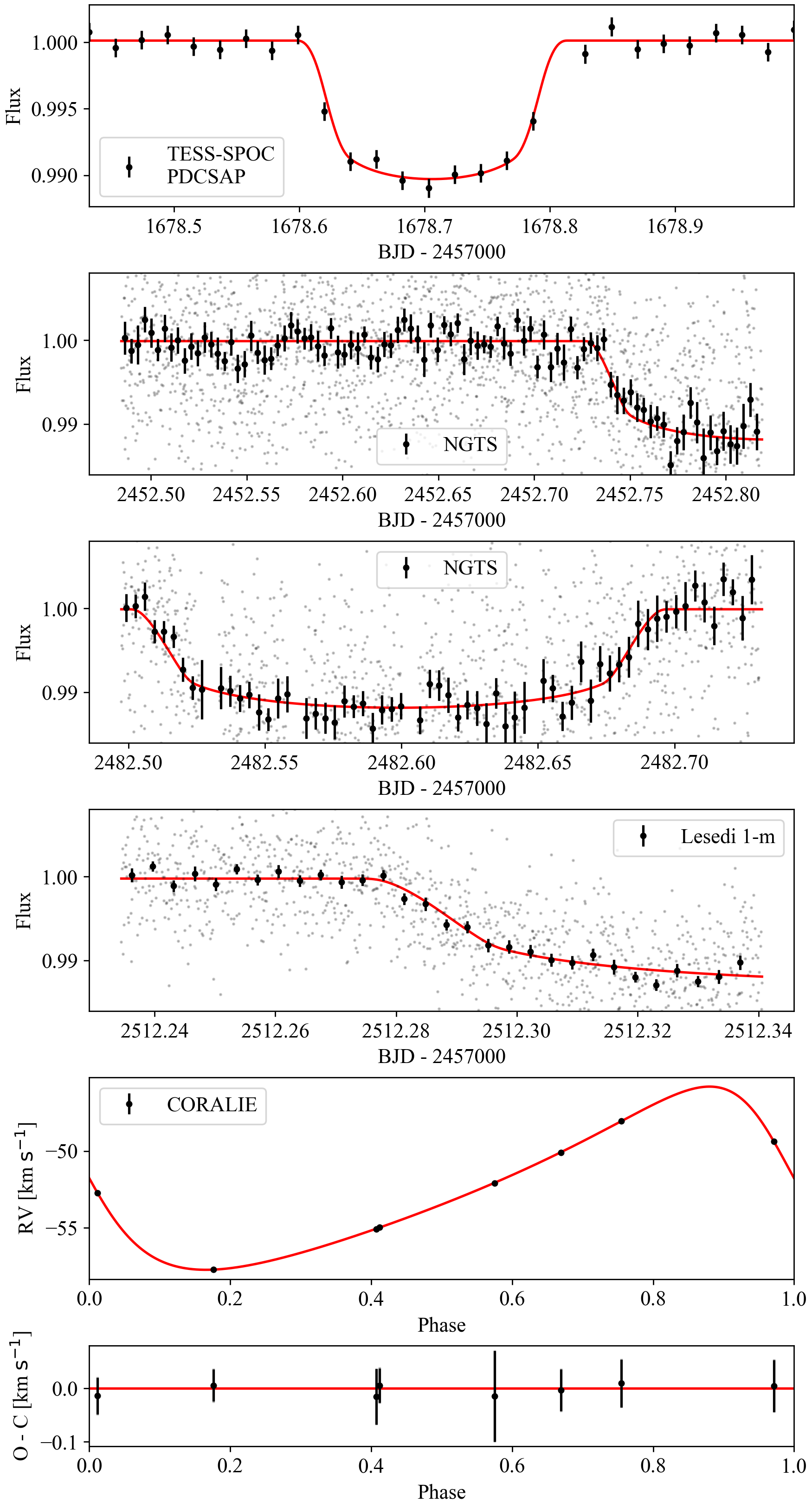}
    \caption{Orbital solution for \system. Transit photomety (black) is shown \tess\ (upper panel), \ngts\ (upper middle panels), and Lesedi (low middle panel) with best fitting models (red). For \ngts\ and Lesedi, we show the 5-minute binned light curve. lower middle panel - \coralie\ radial velocity measurements (black) with best-fitting model (red); Lower panel – fit residuals. }
    \label{fig:fig4}
\end{figure}

\subsection{The \system\ system}

\begin{table}
\caption{Derived and physical properties of the \system\ system. Asymmetric errors are reported in brackets and correspond to the difference between the median and the 16$^{th}$ (lower value) and 84$^{th}$ (upper value) percentile.}              
\label{tab:derived_parameters}      
\centering   
\begin{tabular}{l c}          
\hline\hline                        
Parameter & value\\
\hline 
width [hr] & $5.87 \pm _{0.52}^{0.38}$ \\
$\rho_{\rm A}$ [$\rho_{\odot}$] & $1.73 \pm _{0.24}^{0.25}$ \\
$\log g_{\rm B}$ [dex] & $5.20 \pm _{0.06}^{0.05}$ \\ 
$M_{\rm B}$ [$M_{\odot}$] & $0.0900 \pm _{0.0019}^{0.0018}$ \\
$M_{\rm B}$ [\mjup] & $96.2 \pm _{2.0}^{1.9}$ \\
$R_{\rm B}$ [$R_{\odot}$] & $0.1171 \pm _{0.0023}^{0.0024}$ \\
$R_{\rm B}$ [\rjup] & $1.14 \pm _{0.02}^{0.02}$ \\
$e$ & $0.366 \pm _{0.003}^{0.003}$ \\
$\omega$ [$^{\circ}$] & $68 \pm _{40}^{41}$ \\
$a$ periastron [au] & $0.118 \pm _{0.006}^{0.006}$ \\
$a$ apastron [au] & $0.255 \pm _{0.013}^{0.012}$ \\
\hline
\end{tabular}
\end{table}

Spectral analysis reveals that \systemA\ is richer in metals than the Sun and has spectral type G2V. Gravity-sensitive Mg III and Na II lines appear appear consistent with a star on the main sequence. The transiting companion is a low mass star with an estimated spectral type M7 based on mass and radius measurements. It is expected to be fully convective and we assume any magnetic field will be sustained with a mechanism like the $\alpha^2$-dynamo \citep{2006A&A...446.1027C}. The M-dwarf is close to the best-fitting isochrone (Figure \ref{fig:fig4}) compared to similarly measured objects but we do find a marginally significant inflation (1.8-$\sigma$). In the following sections we discuss interesting aspects of the \system\ system. 

\subsection{Possible inflation of the M-dwarf companion}\label{sec:discuss:inflation}

\begin{figure}
    \centering
    \includegraphics[width=0.47\textwidth]{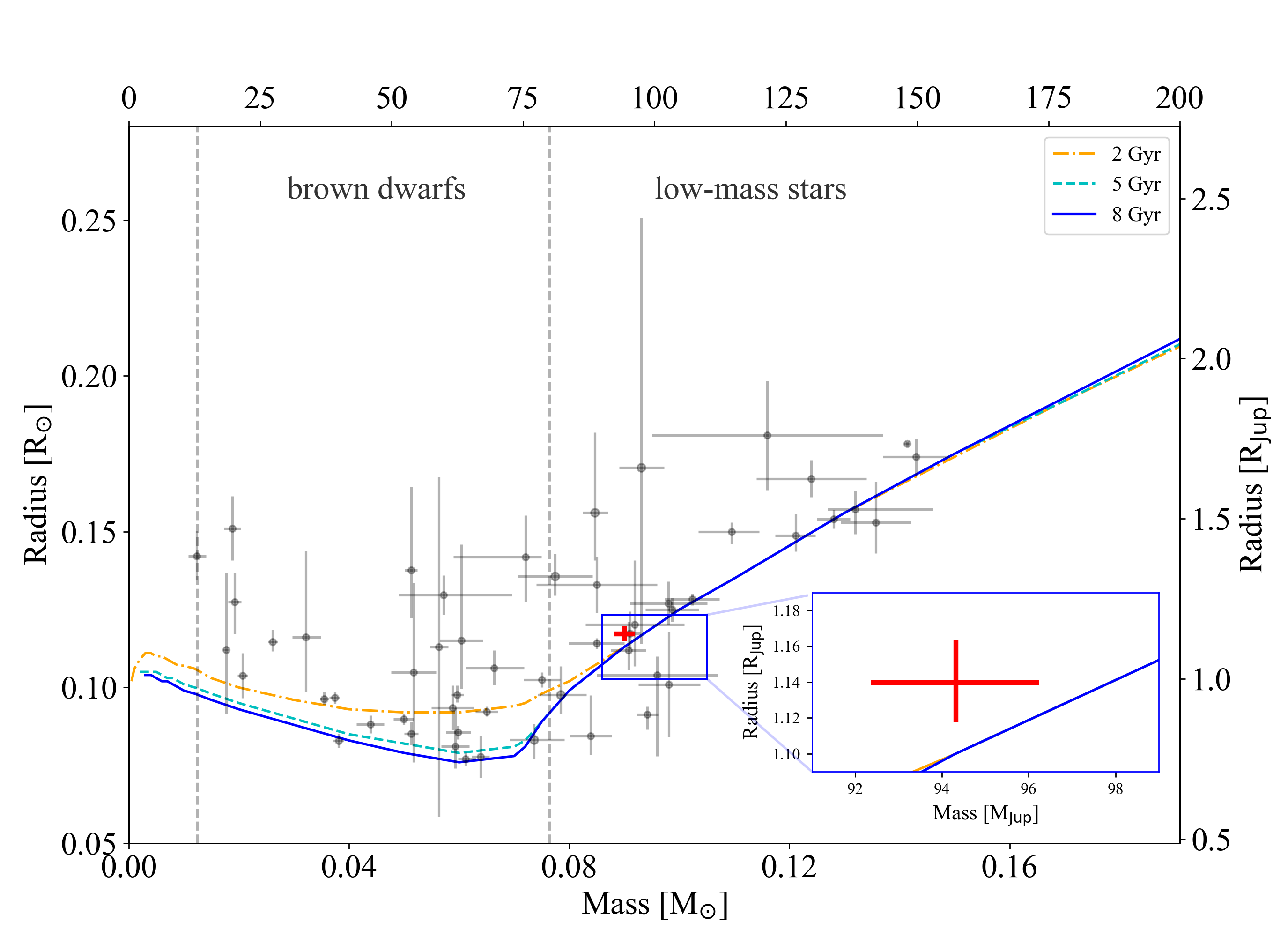}
    \caption{Mass-radius diagram for 54 brown dwarfs and low-mass stars (black) presented in \protect\citep{2021A&A...652A.127G}. Gray vertical lines mark the approximate locations of the planet/brown dwarf boundary ($\sim 13 \, \rm M_{Jup}$) and the brown dwarf/M-dwarf boundary ($\sim 80 \, \rm M_{Jup}$). We show the 5\,Gyr isochrone (cyan) from \citet{2002A&A...382..563B,2015A&A...577A..42B} and mark \systemB\ in red. An inset axis highlights \systemB\ with respect to the best fitting isochrone.}
    \label{fig:mass_radius}
\end{figure}

In Figure \ref{fig:mass_radius} we show the mass and radius of \systemB\ amongst recently measured eclipsing brown dwarfs and late M-dwarfs. \systemB\ is relatively close to stellar models but its measured inflation is marginally significant (1.8-$\sigma$). We assume a coeval formation of the \system\ system around $5 \pm 3$\,Gyr ago. The mass of \systemB\ is where 2-8\,Gyr stellar models show little difference in predicted radii and so age has little bearing on measured inflation. 

The measured inflation may be statistical but there is a possibility it is indeed real. Measuring the radius of \systemB\ is dependent on stellar models used to measure the physical properties of \systemA. These in-turn depend on critical input values of a mixing length parameter and helium enhancement.
\citet{2019A&A...626A.119G} used five EBLM systems to measure an additional 3-5\,\% uncertainty in the mass of the host star when accounting for uncertainties in mixing length parameter and helium enhancement.
The sample used by \citet{2019A&A...626A.119G} consisted of hotter F-type stars and were compared to models from the \textsc{garstec} stellar evolution code \citep{2008Ap&SS.316...99W}. 
Nevertheless
a 3.7\,\% increase in mass would mean \systemB\  was consistent with stellar models.

In contrast to the mass and radius, the surface gravity of \systemB\ is determined entirely from fitted parameters. We find the value $\log g_{\rm B} = 5.20_{(6)}^{(5)}$\,dex is slightly below the expected value of 5.29\,dex from the 5\,Gyr isochrone. 
This result provides evidence for modest inflation independent of models for \systemA, although we note that these values are still consistent within 2-$\sigma$.

We are confident that \systemB\ does not exhibit an enhanced dynamo due to tidal interaction, but it may still be significantly 
spotted. The luminosity ratio between the two components ($\sim 10^{-4}$) is such that we are unable to detect photometric modulation from \systemB\ but it may have a non-negligible spot coverage which could account for the small measured inflation \citep{2005ApJ...631.1120L,2013ApJ...779..183F}. 

\subsection{The hydrogen burning limit}

Brown dwarfs are sub-stellar objects residing between giant planets ($\sim13$\,\mjup) and low-mass main-sequence stars ($\sim80$\,\mjup), with the upper boundary defined by the mass required for stable thermonuclear fusion of hydrogen. During their first few million years, both M dwarfs and brown dwarfs produce energy by fusing deuterium, with their cores contracting and heating up. Fusion of hydrogen via the proton-proton chain requires a sufficiently high pressure that brown dwarfs never reach due in-part to their core density providing a sufficient restoring force with electron degeneracy pressure. Ultimately, M-dwarfs go on to fuse hydrogen for billions of years \citep{1998A&A...337..403B} compared to brown dwarfs that exhaust their comparatively sparse deuterium supply in a few million years before cooling and shrinking \citep{2011ApJ...727...57S}.

The exact transition between brown dwarfs and M dwarfs depends on a number of initial formation conditions including the size of the initial protostar, metal and deuterium abundances, stellar opacity, and the convective efficiency of the outer layers \citep{1997A&A...327.1039C,2002A&A...382..563B}. The generally adopted boundary is $\sim 80$\,\mjup \ \citep[e.g. ][]{2000PASP..112..137M,2006ApJ...640.1051G} which is a median between an array of model predictions spanning 73.3 - 96.4 \mjup\ \citep[see ][ and references therein]{2018AAS...23134918D}. \systemB\ is near the top of this range ($96.2 \pm 0.2$\,\mjup) suggesting that it could reside within the brown dwarf transition. However, its consistency with the 5\,Gyr isochrone suggests it is indeed stellar in nature.

\subsection{Orbital dynamics}\label{sec:discuss:tidal}

The great advantage of systems like \system\ is that their wide separations result in little tidal interaction between components. M-dwarfs characterised in these systems are therefore more akin to isolated field M-dwarfs and can be more robustly compared to stellar models. 
The significant eccentricity of the \system\ system results in an orbital separation of $0.255$\,au at apastron and $0.118$\,au at periastron. 
For long period binaries like \system, we expect tidal circularisation to be exceptionally weak when accounting for the binary mass ratio, $q$. Work by \citet{1997A&A...318..187C} determined a semi-empirical calibration joining the physical parameters of the binary system with critical circularisation and synchronisation timescales for those with convective and radiative envelopes (see their Eqns 15-18). These relations suggest $\tau_{\rm circ}$ is many times greater than the 
age
of the \system\ system and has played a negligible role in any orbital evolution. Additionally, it is possible for a tertiary companion to excite the eccentricity to larger values despite a low primordial eccentricity \citep{1979A&A....77..145M}. We find no evidence of a stellar tertiary companion in either the photometric (transit timing variations) or spectroscopic (radial velocity residuals) datasets and find this scenario unlikely for the \system\ system. It is possible that efficiency of tidal circularisation may have been larger during the pre-main-sequence phase when the host star would have been much larger \citep{1989A&A...220..112Z,1989A&A...223..112Z}. However, this increase would have been marginal given the orbital separation and the expected radii of both stars during the pre-main-sequence.

It is of interest to understand the rotation and stellar inclination of \systemA. One could argue for mutual stellar and orbital inclination as they are relics of the angular momentum from a common primordial cloud. However, their large orbital separation could have resulted in quasi-local formation and dynamical interactions which may not have preserved the alignment between rotation and orbital inclinations. 
If the latter is the case, the timescale of aligning the orbital and spin axis is on the same order as tidal synchronisation
which far exceeds the lifetime of this system \citep{1997A&A...318..187C}. 
Only 4 systems within the EBLM project include measurements of spin-orbit misalignment: WASP-30 and EBLM J1219-39 \citep{2013A&A...549A..18T}; EBLM J0218-31 \citep{2019A&A...626A.119G}; and EBLM J0608-59/TOI-1338 \citep{2020MNRAS.497.1627K} with all suggesting coplanar stellar rotation and orbital axes. The orbital periods of these systems are shorter than \system\ and more likely to be affected by tides. Therefore, it would be of interest to measure the spin-orbit misalignment for \system\ and see if it is consistent with those of shorter orbital periods.

\subsection{Rotational modulation}

The measured rotation period from \ngts\ photometry ($P_{\rm rot}$ = 22.53\,days) corresponds to a surface rotation of $\sim 2.6\, \rm km\,\rm s^{-1}$ which is consistent the spectroscopic value of $V\sin i = 2.5 \pm 0.8\, \rm km\,\rm s^{-1}$. This suggests that the rotation axis is broadly aligned with the orbital axis. We expected mutual stellar and orbital inclinations but past dynamical interactions may have misaligned the two. This can be confirmed with future measurements of the Rossiter–McLaughlin effect for the \system\ system.

\subsection{Secondary eclipse}\label{discuss:secondary}

The orbital dynamics of \system\ indicates a secondary eclipse centred at phase 0.705 (0.7025 - 0.7075) which both \tess\ and \ngts\ datasets cover. We calculate the expected secondary eclipse depth by interpolating \textsc{phoenix} model spectra \citep{2013A&A...553A...6H} for the each component in the \system\ system. For \systemA, we use values of \teff\ and \logg\ from Table \ref{tab:systemAparameters}. For \systemB, we use \logg$_{\rm B}$ from Table \ref{tab:derived_parameters} and use empirical calibrations\footnote{www.pas.rochester.edu/~emamajek/EEM\_dwarf\_UBVIJHK\_colors\_Teff.txt; accessed 2021 Dec 29} to estimate \teff$_{\rm B}$=2680\,K. Transmission filters\footnote{from svo2.cab.inta-csic.es;  accessed 2021 Dec 29} for \ngts\ and \tess\ were used to calculate a secondary eclipse depth of 108\,ppm and 267\,ppm respectively. This is significantly below the noise profile of both \tess\ and \ngts\ datasets and we do not claim any detection of a secondary eclipse. This will be significantly deeper in the infrared where the luminosity ration between \systemA\ and B becomes less extreme. For 2MASS filters $J$, $H$, and $K_{\rm s}$ we calculate secondary eclipse depths of 1.1\,ppt, 1.4\,ppt, and 1.9\,ppt respectively. This is within the capabilities of modern ground-based infrared telescopes and would provide an measurement of the stellar effective temperature for \systemB. EBLM systems with measured secondary eclipses have revealed M-dwarfs with effective temperatures in excess of predicted by stellar models \citep[e.g. J0113+31;][]{2014A&A...572A..50G} and it would be of interest to see if \systemB\ is similar.

\section{Conclusion}

\system\ is a 
long-period
EBLM system with a very low-mass secondary star close to the hydrogen burning limit. Tidal effects on both components are negligible due to the wide orbital separation. 

The low-mass companion \systemB\ was initially identified through a single-transit event in \tess\ full-frame lightcurves from Sector 13. The transit depth and width was consistent with a Jovian planet and so we commenced a ground-based spectroscopic and photometric campaign to recover the orbital period. A total of 8 \coralie\ radial velocities (Table \ref{tab:radial_velocities}) first provided an approximate spectroscopic orbit followed by 2 transits with \ngts\ and 1 transit with Lesedi which confirmed an orbital period of 29.77381\,d. Spectroscopic observations were used to measure physical and stellar atmospheric parameters of \systemA. They confirmed a spectral type G2 with mass, temperature, radius, and age similar to the Sun (Table \ref{tab:systemAparameters} and Figure \ref{fig:primary_star}). 

Joint analysis of photometric and spectroscopic datasets (Figure \ref{fig:fig4}, Tables \ref{tab:fitted_parameters} \& \ref{tab:derived_parameters}) revealed \systemB\ to be a late M-dwarf ($M_{\rm B}$ = $96.2 \pm _{2.0}^{1.9}$ \mjup, $R_{\rm B}$ = $1.14 \pm _{0.02}^{0.02}$ \rjup) near the hydrogen burning limit ($\sim 80$\,\mjup). The mass and radius of \systemB\ is 
closer to stellar models 
than many
other M-dwarfs from the literature (mostly in much closer binaries). However, we do find a marginally significant inflation (1.8-$\sigma$) which might be statistical or may be a real offset. We measure a moderately high eccentricity of the \system\ system ($e = 0.366 \pm 0.003$) which likely remains from formation due to a large orbital separation diminishing any tidal influence between components. With \ngts\ we measure a likely spot modulation indicating a rotational period of 22.53\, days which is consistent with the projected rotation measured from spectroscopic analysis. 

Most EBLMs with precise measurements of physical parameters have orbital periods below $\sim10\,\rm d$ and are subjected to tidal interactions which complicate discussions of systematic inflation. Longer period systems tend to be free of this and are more akin to field M-dwarfs, which are the subject of intense surveys for small transiting exoplanets by \tess\ and other instruments. 
An increasing number of precisely measured systems like \system\ will allow us to test models of stellar evolution for the smallest main-sequence stars and better understand the planets we find around them.

\section*{Acknowledgements}

The \ngts\ facility is operated by the consortium institutes with support from the UK Science and Technology Facilities Council (STFC) under projects ST/M001962/1 and ST/S002642/1. 
We acknowledge the use of public TESS data from pipelines at the TESS Science Office and at the TESS Science Processing Operations Centre.
This paper includes data collected with the TESS mission, obtained from the MAST data archive at the Space Telescope Science Institute (STScI). Funding for the TESS mission is provided by the NASA Explorer Program. STScI is operated by the Association of Universities for Research in Astronomy, Inc., under NASA contract NAS 5–26555.
Based on observations made with ESO Telescopes at the La Silla Paranal Observatory under programme IDs $0104.C-0413$ (PI RB), $0104.C-0588$ (PI FB), Opticon:2019A/037 (PI DB), and CNTAC: $0104.A-9012$ (PI JIV).
This paper uses observations made at the South African Astronomical Observatory (SAAO).
The contributions at the University of Warwick by PJW, RGW, DRA, and SG have been supported by STFC through consolidated grants ST/L000733/1 and ST/P000495/1.
Contributions at the University of Geneva by SUl, NG, ML, FB, LM, and SUd were carried out within the framework of the National Centre for Competence in Research ``PlanetS'' supported by the Swiss National Science Foundation (SNSF).
ML acknowledges support of the Swiss National Science Foundation under grant number PCEFP2194576.
This research has made use of NASA's Astrophysics Data System Bibliographic Services and the SIMBAD database, operated at CDS, Strasbourg, France. This research made use of Astropy,\footnote{www.astropy.org} a community-developed core Python package for Astronomy \citep{2018AJ....156..123A}.
MNG acknowledges support from the European Space Agency (ESA) as an ESA Research Fellow.
JSJ acknowledges support by FONDECYT grant 1201371 and partial support from the ANID Basal project FB210003.
The work of HPO has been carried out within the framework of the NCCR PlanetS supported by the Swiss National Science Foundation.
EG gratefully acknowledges support from the David and Claudia Harding Foundation in the form of a Winton Exoplanet Fellowship.
\section*{Data Availability}

The \tess\ SPOC data for \systemt\ is publicly available on the Mikulski Archive for Space Telescopes (MAST). Reduced \coralie\ spectra, derived measurements of radial velocities, and full photometric datasets from \ngts\ and Lesedi will be available from the VizieR archive server hosted by the Universit\'{e} de Strasbourg.\footnote{cdsarc.u-strasbg.fr}.



\bibliographystyle{mnras}
\bibliography{paper} 

\begin{thebibliography}{}
\makeatletter
\relax
\def\mn@urlcharsother{\let\do\@makeother \do\$\do\&\do\#\do\^\do\_\do\%\do\~}
\def\mn@doi{\begingroup\mn@urlcharsother \@ifnextchar [ {\mn@doi@}
  {\mn@doi@[]}}
\def\mn@doi@[#1]#2{\def\@tempa{#1}\ifx\@tempa\@empty \href
  {http://dx.doi.org/#2} {doi:#2}\else \href {http://dx.doi.org/#2} {#1}\fi
  \endgroup}
\def\mn@eprint#1#2{\mn@eprint@#1:#2::\@nil}
\def\mn@eprint@arXiv#1{\href {http://arxiv.org/abs/#1} {{\tt arXiv:#1}}}
\def\mn@eprint@dblp#1{\href {http://dblp.uni-trier.de/rec/bibtex/#1.xml}
  {dblp:#1}}
\def\mn@eprint@#1:#2:#3:#4\@nil{\def\@tempa {#1}\def\@tempb {#2}\def\@tempc
  {#3}\ifx \@tempc \@empty \let \@tempc \@tempb \let \@tempb \@tempa \fi \ifx
  \@tempb \@empty \def\@tempb {arXiv}\fi \@ifundefined
  {mn@eprint@\@tempb}{\@tempb:\@tempc}{\expandafter \expandafter \csname
  mn@eprint@\@tempb\endcsname \expandafter{\@tempc}}}

\bibitem[\protect\citeauthoryear{{Asplund}, {Grevesse}, {Sauval}  \&
  {Scott}}{{Asplund} et~al.}{2009}]{2009ARA&A..47..481A}
{Asplund} M.,  {Grevesse} N.,  {Sauval} A.~J.,   {Scott} P.,  2009, \mn@doi
  [\araa] {10.1146/annurev.astro.46.060407.145222}, \href
  {https://ui.adsabs.harvard.edu/abs/2009ARA&A..47..481A} {47, 481}

\bibitem[\protect\citeauthoryear{{Astropy Collaboration} et~al.,}{{Astropy
  Collaboration} et~al.}{2018}]{2018AJ....156..123A}
{Astropy Collaboration} et~al., 2018, \mn@doi [\aj] {10.3847/1538-3881/aabc4f},
  \href {https://ui.adsabs.harvard.edu/abs/2018AJ....156..123A} {156, 123}

\bibitem[\protect\citeauthoryear{{Baraffe}, {Chabrier}, {Allard}  \&
  {Hauschildt}}{{Baraffe} et~al.}{1998}]{1998A&A...337..403B}
{Baraffe} I.,  {Chabrier} G.,  {Allard} F.,   {Hauschildt} P.~H.,  1998, \aap,
  \href {https://ui.adsabs.harvard.edu/abs/1998A&A...337..403B} {337, 403}

\bibitem[\protect\citeauthoryear{{Baraffe}, {Chabrier}, {Allard}  \&
  {Hauschildt}}{{Baraffe} et~al.}{2002}]{2002A&A...382..563B}
{Baraffe} I.,  {Chabrier} G.,  {Allard} F.,   {Hauschildt} P.~H.,  2002,
  \mn@doi [\aap] {10.1051/0004-6361:20011638}, \href
  {https://ui.adsabs.harvard.edu/abs/2002A&A...382..563B} {382, 563}

\bibitem[\protect\citeauthoryear{{Baraffe}, {Homeier}, {Allard}  \&
  {Chabrier}}{{Baraffe} et~al.}{2015}]{2015A&A...577A..42B}
{Baraffe} I.,  {Homeier} D.,  {Allard} F.,   {Chabrier} G.,  2015, \mn@doi
  [\aap] {10.1051/0004-6361/201425481}, \href
  {https://ui.adsabs.harvard.edu/abs/2015A&A...577A..42B} {577, A42}

\bibitem[\protect\citeauthoryear{{Berger} et~al.,}{{Berger}
  et~al.}{2006}]{2006ApJ...644..475B}
{Berger} D.~H.,  et~al., 2006, \mn@doi [\apj] {10.1086/503318}, \href
  {http://adsabs.harvard.edu/abs/2006ApJ...644..475B} {644, 475}

\bibitem[\protect\citeauthoryear{{Burt} et~al.,}{{Burt}
  et~al.}{2020}]{2020AJ....160..153B}
{Burt} J.~A.,  et~al., 2020, \mn@doi [\aj] {10.3847/1538-3881/abac0c}, \href
  {https://ui.adsabs.harvard.edu/abs/2020AJ....160..153B} {160, 153}

\bibitem[\protect\citeauthoryear{{Caldwell} et~al.,}{{Caldwell}
  et~al.}{2020}]{2020RNAAS...4..201C}
{Caldwell} D.~A.,  et~al., 2020, \mn@doi [Research Notes of the American
  Astronomical Society] {10.3847/2515-5172/abc9b3}, \href
  {https://ui.adsabs.harvard.edu/abs/2020RNAAS...4..201C} {4, 201}

\bibitem[\protect\citeauthoryear{{Chabrier} \& {Baraffe}}{{Chabrier} \&
  {Baraffe}}{1997}]{1997A&A...327.1039C}
{Chabrier} G.,  {Baraffe} I.,  1997, \aap, \href
  {https://ui.adsabs.harvard.edu/abs/1997A&A...327.1039C} {327, 1039}

\bibitem[\protect\citeauthoryear{{Chabrier} \& {K{\"u}ker}}{{Chabrier} \&
  {K{\"u}ker}}{2006}]{2006A&A...446.1027C}
{Chabrier} G.,  {K{\"u}ker} M.,  2006, \mn@doi [\aap]
  {10.1051/0004-6361:20042475}, \href
  {https://ui.adsabs.harvard.edu/abs/2006A&A...446.1027C} {446, 1027}

\bibitem[\protect\citeauthoryear{{Chabrier}, {Gallardo}  \&
  {Baraffe}}{{Chabrier} et~al.}{2007}]{2007AA...472L..17C}
{Chabrier} G.,  {Gallardo} J.,   {Baraffe} I.,  2007, \mn@doi [\aap]
  {10.1051/0004-6361:20077702}, \href
  {https://ui.adsabs.harvard.edu/abs/2007A&A...472L..17C} {472, L17}

\bibitem[\protect\citeauthoryear{{Choi}, {Dotter}, {Conroy}, {Cantiello},
  {Paxton}  \& {Johnson}}{{Choi} et~al.}{2016}]{2016ApJ...823..102C}
{Choi} J.,  {Dotter} A.,  {Conroy} C.,  {Cantiello} M.,  {Paxton} B.,
  {Johnson} B.~D.,  2016, \mn@doi [\apj] {10.3847/0004-637X/823/2/102}, \href
  {https://ui.adsabs.harvard.edu/abs/2016ApJ...823..102C} {823, 102}

\bibitem[\protect\citeauthoryear{{Claret} \& {Cunha}}{{Claret} \&
  {Cunha}}{1997}]{1997A&A...318..187C}
{Claret} A.,  {Cunha} N.~C.~S.,  1997, \aap, \href
  {https://ui.adsabs.harvard.edu/abs/1997A&A...318..187C} {318, 187}

\bibitem[\protect\citeauthoryear{{Coppejans} et~al.,}{{Coppejans}
  et~al.}{2013}]{2013PASP..125..976C}
{Coppejans} R.,  et~al., 2013, \mn@doi [\pasp] {10.1086/672156}, \href
  {https://ui.adsabs.harvard.edu/abs/2013PASP..125..976C} {125, 976}

\bibitem[\protect\citeauthoryear{{Dieterich}, {Henry}, {Jao}, {Washington},
  {Silverstein}, {Winters}  \& {RECONS}}{{Dieterich}
  et~al.}{2018}]{2018AAS...23134918D}
{Dieterich} S.,  {Henry} T.,  {Jao} W.~C.,  {Washington} R.,  {Silverstein} M.,
   {Winters} J.,   {RECONS} 2018, in American Astronomical Society Meeting
  Abstracts \#231. p. 349.18

\bibitem[\protect\citeauthoryear{{Dotter}}{{Dotter}}{2016}]{2016ApJS..222....8D}
{Dotter} A.,  2016, \mn@doi [\apjs] {10.3847/0067-0049/222/1/8}, \href
  {https://ui.adsabs.harvard.edu/abs/2016ApJS..222....8D} {222, 8}

\bibitem[\protect\citeauthoryear{{Doyle}}{{Doyle}}{2015}]{2015PhDT........16D}
{Doyle} A.~P.,  2015, PhD thesis, Keele University

\bibitem[\protect\citeauthoryear{{Feiden} \& {Chaboyer}}{{Feiden} \&
  {Chaboyer}}{2013a}]{2013EAS....64..127F}
{Feiden} G.~A.,  {Chaboyer} B.,  2013a, in {Pavlovski} K.,  {Tkachenko} A.,
  {Torres} G.,  eds,  EAS Publications Series Vol. 64, EAS Publications Series.
  pp 127--130 (\mn@eprint {arXiv} {1310.8567}), \mn@doi{10.1051/eas/1364017}

\bibitem[\protect\citeauthoryear{{Feiden} \& {Chaboyer}}{{Feiden} \&
  {Chaboyer}}{2013b}]{2013ApJ...779..183F}
{Feiden} G.~A.,  {Chaboyer} B.,  2013b, \mn@doi [\apj]
  {10.1088/0004-637X/779/2/183}, \href
  {http://adsabs.harvard.edu/abs/2013ApJ...779..183F} {779, 183}

\bibitem[\protect\citeauthoryear{{Ford}}{{Ford}}{2006}]{2006ApJ...642..505F}
{Ford} E.~B.,  2006, \mn@doi [\apj] {10.1086/500802}, \href
  {https://ui.adsabs.harvard.edu/abs/2006ApJ...642..505F} {642, 505}

\bibitem[\protect\citeauthoryear{{Foreman-Mackey}, {Hogg}, {Lang}  \&
  {Goodman}}{{Foreman-Mackey} et~al.}{2013}]{2013PASP..125..306F}
{Foreman-Mackey} D.,  {Hogg} D.~W.,  {Lang} D.,   {Goodman} J.,  2013, \mn@doi
  [\pasp] {10.1086/670067}, \href
  {https://ui.adsabs.harvard.edu/abs/2013PASP..125..306F} {125, 306}

\bibitem[\protect\citeauthoryear{{Gaia Collaboration} et~al.,}{{Gaia
  Collaboration} et~al.}{2018}]{2018A&A...616A...1G}
{Gaia Collaboration} et~al., 2018, \mn@doi [\aap]
  {10.1051/0004-6361/201833051}, \href
  {https://ui.adsabs.harvard.edu/abs/2018A&A...616A...1G} {616, A1}

\bibitem[\protect\citeauthoryear{{Gill} et~al.,}{{Gill}
  et~al.}{2019}]{2019A&A...626A.119G}
{Gill} S.,  et~al., 2019, \mn@doi [\aap] {10.1051/0004-6361/201833054}, \href
  {https://ui.adsabs.harvard.edu/abs/2019A&A...626A.119G} {626, A119}

\bibitem[\protect\citeauthoryear{{Gill} et~al.,}{{Gill}
  et~al.}{2020a}]{2020MNRAS.491.1548G}
{Gill} S.,  et~al., 2020a, \mn@doi [\mnras] {10.1093/mnras/stz3212}, \href
  {https://ui.adsabs.harvard.edu/abs/2020MNRAS.491.1548G} {491, 1548}

\bibitem[\protect\citeauthoryear{{Gill} et~al.,}{{Gill}
  et~al.}{2020b}]{2020MNRAS.495.2713G}
{Gill} S.,  et~al., 2020b, \mn@doi [\mnras] {10.1093/mnras/staa1248}, \href
  {https://ui.adsabs.harvard.edu/abs/2020MNRAS.495.2713G} {495, 2713}

\bibitem[\protect\citeauthoryear{{Gill} et~al.,}{{Gill}
  et~al.}{2020c}]{2020ApJ...898L..11G}
{Gill} S.,  et~al., 2020c, \mn@doi [\apjl] {10.3847/2041-8213/ab9eb9}, \href
  {https://ui.adsabs.harvard.edu/abs/2020ApJ...898L..11G} {898, L11}

\bibitem[\protect\citeauthoryear{{Gillon} et~al.,}{{Gillon}
  et~al.}{2016}]{2016Natur.533..221G}
{Gillon} M.,  et~al., 2016, \mn@doi [\nat] {10.1038/nature17448}, \href
  {https://ui.adsabs.harvard.edu/abs/2016Natur.533..221G} {533, 221}

\bibitem[\protect\citeauthoryear{{Gillon} et~al.,}{{Gillon}
  et~al.}{2017}]{2017Natur.542..456G}
{Gillon} M.,  et~al., 2017, \mn@doi [\nat] {10.1038/nature21360}, \href
  {https://ui.adsabs.harvard.edu/abs/2017Natur.542..456G} {542, 456}

\bibitem[\protect\citeauthoryear{{G{\'o}mez Maqueo Chew} et~al.,}{{G{\'o}mez
  Maqueo Chew} et~al.}{2014}]{2014A&A...572A..50G}
{G{\'o}mez Maqueo Chew} Y.,  et~al., 2014, \mn@doi [\aap]
  {10.1051/0004-6361/201424265}, \href
  {https://ui.adsabs.harvard.edu/abs/2014A&A...572A..50G} {572, A50}

\bibitem[\protect\citeauthoryear{{Gray}}{{Gray}}{1999}]{1999ascl.soft10002G}
{Gray} R.~O.,  1999, {SPECTRUM: A stellar spectral synthesis program}
  (\mn@eprint {ascl} {9910.002})

\bibitem[\protect\citeauthoryear{{Grether} \& {Lineweaver}}{{Grether} \&
  {Lineweaver}}{2006}]{2006ApJ...640.1051G}
{Grether} D.,  {Lineweaver} C.~H.,  2006, \mn@doi [\apj] {10.1086/500161},
  \href {https://ui.adsabs.harvard.edu/abs/2006ApJ...640.1051G} {640, 1051}

\bibitem[\protect\citeauthoryear{{Grieves} et~al.,}{{Grieves}
  et~al.}{2021}]{2021A&A...652A.127G}
{Grieves} N.,  et~al., 2021, \mn@doi [\aap] {10.1051/0004-6361/202141145},
  \href {https://ui.adsabs.harvard.edu/abs/2021A&A...652A.127G} {652, A127}

\bibitem[\protect\citeauthoryear{{Gustafsson}, {Edvardsson}, {Eriksson},
  {J{\o}rgensen}, {Nordlund}  \& {Plez}}{{Gustafsson}
  et~al.}{2008}]{2008A&A...486..951G}
{Gustafsson} B.,  {Edvardsson} B.,  {Eriksson} K.,  {J{\o}rgensen} U.~G.,
  {Nordlund} {\r{A}}.,   {Plez} B.,  2008, \mn@doi [\aap]
  {10.1051/0004-6361:200809724}, \href
  {https://ui.adsabs.harvard.edu/abs/2008A&A...486..951G} {486, 951}

\bibitem[\protect\citeauthoryear{{Henden}, {Levine}, {Terrell}  \&
  {Welch}}{{Henden} et~al.}{2015}]{2015AAS...22533616H}
{Henden} A.~A.,  {Levine} S.,  {Terrell} D.,   {Welch} D.~L.,  2015, in
  American Astronomical Society Meeting Abstracts \#225. p. 336.16

\bibitem[\protect\citeauthoryear{{Hsu}, {Ford}  \& {Terrien}}{{Hsu}
  et~al.}{2020}]{2020MNRAS.498.2249H}
{Hsu} D.~C.,  {Ford} E.~B.,   {Terrien} R.,  2020, \mn@doi [\mnras]
  {10.1093/mnras/staa2391}, \href
  {https://ui.adsabs.harvard.edu/abs/2020MNRAS.498.2249H} {498, 2249}

\bibitem[\protect\citeauthoryear{{Husser}, {Wende-von Berg}, {Dreizler},
  {Homeier}, {Reiners}, {Barman}  \& {Hauschildt}}{{Husser}
  et~al.}{2013}]{2013A&A...553A...6H}
{Husser} T.~O.,  {Wende-von Berg} S.,  {Dreizler} S.,  {Homeier} D.,  {Reiners}
  A.,  {Barman} T.,   {Hauschildt} P.~H.,  2013, \mn@doi [\aap]
  {10.1051/0004-6361/201219058}, \href
  {https://ui.adsabs.harvard.edu/abs/2013A&A...553A...6H} {553, A6}

\bibitem[\protect\citeauthoryear{{Jenkins} et~al.,}{{Jenkins}
  et~al.}{2016}]{2016SPIE.9913E..3EJ}
{Jenkins} J.~M.,  et~al., 2016, in {Chiozzi} G.,  {Guzman} J.~C.,  eds,
  Society of Photo-Optical Instrumentation Engineers (SPIE) Conference Series
  Vol. 9913, Software and Cyberinfrastructure for Astronomy IV. p. 99133E,
  \mn@doi{10.1117/12.2233418}

\bibitem[\protect\citeauthoryear{{Kraus}, {Tucker}, {Thompson}, {Craine}  \&
  {Hillenbrand}}{{Kraus} et~al.}{2011}]{2011ApJ...728...48K}
{Kraus} A.~L.,  {Tucker} R.~A.,  {Thompson} M.~I.,  {Craine} E.~R.,
  {Hillenbrand} L.~A.,  2011, \mn@doi [\apj] {10.1088/0004-637X/728/1/48},
  \href {https://ui.adsabs.harvard.edu/abs/2011ApJ...728...48K} {728, 48}

\bibitem[\protect\citeauthoryear{{Kunovac Hod{\v{z}}i{\'c}} et~al.,}{{Kunovac
  Hod{\v{z}}i{\'c}} et~al.}{2020}]{2020MNRAS.497.1627K}
{Kunovac Hod{\v{z}}i{\'c}} V.,  et~al., 2020, \mn@doi [\mnras]
  {10.1093/mnras/staa2071}, \href
  {https://ui.adsabs.harvard.edu/abs/2020MNRAS.497.1627K} {497, 1627}

\bibitem[\protect\citeauthoryear{{Leggett}, {Allard}, {Dahn}, {Hauschildt},
  {Kerr}  \& {Rayner}}{{Leggett} et~al.}{2000}]{2000ApJ...535..965L}
{Leggett} S.~K.,  {Allard} F.,  {Dahn} C.,  {Hauschildt} P.~H.,  {Kerr} T.~H.,
   {Rayner} J.,  2000, \mn@doi [\apj] {10.1086/308887}, \href
  {http://adsabs.harvard.edu/abs/2000ApJ...535..965L} {535, 965}

\bibitem[\protect\citeauthoryear{{Lendl} et~al.,}{{Lendl}
  et~al.}{2020}]{2020MNRAS.492.1761L}
{Lendl} M.,  et~al., 2020, \mn@doi [\mnras] {10.1093/mnras/stz3545}, \href
  {https://ui.adsabs.harvard.edu/abs/2020MNRAS.492.1761L} {492, 1761}

\bibitem[\protect\citeauthoryear{{L{\'o}pez-Morales}}{{L{\'o}pez-Morales}}{2007}]{2007ApJ...660..732L}
{L{\'o}pez-Morales} M.,  2007, \mn@doi [\apj] {10.1086/513142}, \href
  {http://adsabs.harvard.edu/abs/2007ApJ...660..732L} {660, 732}

\bibitem[\protect\citeauthoryear{{L{\'o}pez-Morales} \&
  {Ribas}}{{L{\'o}pez-Morales} \& {Ribas}}{2005}]{2005ApJ...631.1120L}
{L{\'o}pez-Morales} M.,  {Ribas} I.,  2005, \mn@doi [\apj] {10.1086/432680},
  \href {http://adsabs.harvard.edu/abs/2005ApJ...631.1120L} {631, 1120}

\bibitem[\protect\citeauthoryear{{Lubin} et~al.,}{{Lubin}
  et~al.}{2017}]{2017ApJ...844..134L}
{Lubin} J.~B.,  et~al., 2017, \mn@doi [\apj] {10.3847/1538-4357/aa7947}, \href
  {https://ui.adsabs.harvard.edu/abs/2017ApJ...844..134L} {844, 134}

\bibitem[\protect\citeauthoryear{{Marcy} \& {Butler}}{{Marcy} \&
  {Butler}}{2000}]{2000PASP..112..137M}
{Marcy} G.~W.,  {Butler} R.~P.,  2000, \mn@doi [\pasp] {10.1086/316516}, \href
  {https://ui.adsabs.harvard.edu/abs/2000PASP..112..137M} {112, 137}

\bibitem[\protect\citeauthoryear{{Maxted}}{{Maxted}}{2018}]{2018A&A...616A..39M}
{Maxted} P.~F.~L.,  2018, \mn@doi [\aap] {10.1051/0004-6361/201832944}, \href
  {https://ui.adsabs.harvard.edu/abs/2018A&A...616A..39M} {616, A39}

\bibitem[\protect\citeauthoryear{{Maxted} \& {Gill}}{{Maxted} \&
  {Gill}}{2019}]{2019A&A...622A..33M}
{Maxted} P.~F.~L.,  {Gill} S.,  2019, \mn@doi [\aap]
  {10.1051/0004-6361/201834563}, \href
  {https://ui.adsabs.harvard.edu/abs/2019A&A...622A..33M} {622, A33}

\bibitem[\protect\citeauthoryear{{Mazeh} \& {Shaham}}{{Mazeh} \&
  {Shaham}}{1979}]{1979A&A....77..145M}
{Mazeh} T.,  {Shaham} J.,  1979, \aap, \href
  {https://ui.adsabs.harvard.edu/abs/1979A&A....77..145M} {77, 145}

\bibitem[\protect\citeauthoryear{{Morris}, {Twicken}, {Smith}, {Clarke},
  {Jenkins}, {Bryson}, {Girouard}  \& {Klaus}}{{Morris}
  et~al.}{2017}]{2017ksci.rept....6M}
{Morris} R.~L.,  {Twicken} J.~D.,  {Smith} J.~C.,  {Clarke} B.~D.,  {Jenkins}
  J.~M.,  {Bryson} S.~T.,  {Girouard} F.,   {Klaus} T.~C.,  2017, {Kepler Data
  Processing Handbook: Photometric Analysis}, Kepler Science Document
  KSCI-19081-002

\bibitem[\protect\citeauthoryear{{Nutzman} \& {Charbonneau}}{{Nutzman} \&
  {Charbonneau}}{2008}]{2008PASP..120..317N}
{Nutzman} P.,  {Charbonneau} D.,  2008, \mn@doi [\pasp] {10.1086/533420}, \href
  {https://ui.adsabs.harvard.edu/abs/2008PASP..120..317N} {120, 317}

\bibitem[\protect\citeauthoryear{{Parsons} et~al.,}{{Parsons}
  et~al.}{2018}]{2018MNRAS.481.1083P}
{Parsons} S.~G.,  et~al., 2018, \mn@doi [\mnras] {10.1093/mnras/sty2345}, \href
  {https://ui.adsabs.harvard.edu/abs/2018MNRAS.481.1083P} {481, 1083}

\bibitem[\protect\citeauthoryear{{Pollacco} et~al.,}{{Pollacco}
  et~al.}{2006}]{2006PASP..118.1407P}
{Pollacco} D.~L.,  et~al., 2006, \mn@doi [\pasp] {10.1086/508556}, \href
  {https://ui.adsabs.harvard.edu/abs/2006PASP..118.1407P} {118, 1407}

\bibitem[\protect\citeauthoryear{{Queloz} et~al.,}{{Queloz}
  et~al.}{2001}]{2001A&A...379..279Q}
{Queloz} D.,  et~al., 2001, \mn@doi [\aap] {10.1051/0004-6361:20011308}, \href
  {http://adsabs.harvard.edu/abs/2001A%26A...379..279Q} {379, 279}

\bibitem[\protect\citeauthoryear{{Ricker} et~al.,}{{Ricker}
  et~al.}{2015}]{2015JATIS...1a4003R}
{Ricker} G.~R.,  et~al., 2015, \mn@doi [Journal of Astronomical Telescopes,
  Instruments, and Systems] {10.1117/1.JATIS.1.1.014003}, \href
  {https://ui.adsabs.harvard.edu/abs/2015JATIS...1a4003R} {1, 014003}

\bibitem[\protect\citeauthoryear{{Seager} \& {Mall{\'e}n-Ornelas}}{{Seager} \&
  {Mall{\'e}n-Ornelas}}{2003}]{2003ApJ...585.1038S}
{Seager} S.,  {Mall{\'e}n-Ornelas} G.,  2003, \mn@doi [\apj] {10.1086/346105},
  \href {https://ui.adsabs.harvard.edu/abs/2003ApJ...585.1038S} {585, 1038}

\bibitem[\protect\citeauthoryear{{Sebastian} et~al.,}{{Sebastian}
  et~al.}{2021}]{2021A&A...645A.100S}
{Sebastian} D.,  et~al., 2021, \mn@doi [\aap] {10.1051/0004-6361/202038827},
  \href {https://ui.adsabs.harvard.edu/abs/2021A&A...645A.100S} {645, A100}

\bibitem[\protect\citeauthoryear{{Skrutskie} et~al.,}{{Skrutskie}
  et~al.}{2006}]{2006AJ....131.1163S}
{Skrutskie} M.~F.,  et~al., 2006, \mn@doi [\aj] {10.1086/498708}, \href
  {https://ui.adsabs.harvard.edu/abs/2006AJ....131.1163S} {131, 1163}

\bibitem[\protect\citeauthoryear{{Southworth}, {Wheatley}  \&
  {Sams}}{{Southworth} et~al.}{2007}]{2007MNRAS.379L..11S}
{Southworth} J.,  {Wheatley} P.~J.,   {Sams} G.,  2007, \mn@doi [\mnras]
  {10.1111/j.1745-3933.2007.00324.x}, \href
  {https://ui.adsabs.harvard.edu/abs/2007MNRAS.379L..11S} {379, L11}

\bibitem[\protect\citeauthoryear{{Spiegel}, {Burrows}  \& {Milsom}}{{Spiegel}
  et~al.}{2011}]{2011ApJ...727...57S}
{Spiegel} D.~S.,  {Burrows} A.,   {Milsom} J.~A.,  2011, \mn@doi [\apj]
  {10.1088/0004-637X/727/1/57}, \href
  {https://ui.adsabs.harvard.edu/abs/2011ApJ...727...57S} {727, 57}

\bibitem[\protect\citeauthoryear{{Stassun} et~al.,}{{Stassun}
  et~al.}{2018}]{2018AJ....156..102S}
{Stassun} K.~G.,  et~al., 2018, \mn@doi [\aj] {10.3847/1538-3881/aad050}, \href
  {https://ui.adsabs.harvard.edu/abs/2018AJ....156..102S} {156, 102}

\bibitem[\protect\citeauthoryear{{Triaud} et~al.,}{{Triaud}
  et~al.}{2013}]{2013A&A...549A..18T}
{Triaud} A.~H.~M.~J.,  et~al., 2013, \mn@doi [\aap]
  {10.1051/0004-6361/201219643}, \href
  {https://ui.adsabs.harvard.edu/abs/2013A&A...549A..18T} {549, A18}

\bibitem[\protect\citeauthoryear{{Triaud} et~al.,}{{Triaud}
  et~al.}{2017}]{2017A&A...608A.129T}
{Triaud} A. H.~M.~J.,  et~al., 2017, \mn@doi [\aap]
  {10.1051/0004-6361/201730993}, \href
  {https://ui.adsabs.harvard.edu/abs/2017A&A...608A.129T} {608, A129}

\bibitem[\protect\citeauthoryear{{Twicken}, {Clarke}, {Bryson}, {Tenenbaum},
  {Wu}, {Jenkins}, {Girouard}  \& {Klaus}}{{Twicken}
  et~al.}{2010}]{2010SPIE.7740E..23T}
{Twicken} J.~D.,  {Clarke} B.~D.,  {Bryson} S.~T.,  {Tenenbaum} P.,  {Wu} H.,
  {Jenkins} J.~M.,  {Girouard} F.,   {Klaus} T.~C.,  2010, in {Radziwill}
  N.~M.,  {Bridger} A.,  eds,  Society of Photo-Optical Instrumentation
  Engineers (SPIE) Conference Series Vol. 7740, Software and
  Cyberinfrastructure for Astronomy. p. 774023, \mn@doi{10.1117/12.856790}

\bibitem[\protect\citeauthoryear{{Van Eylen} \& {Albrecht}}{{Van Eylen} \&
  {Albrecht}}{2015}]{2015ApJ...808..126V}
{Van Eylen} V.,  {Albrecht} S.,  2015, \mn@doi [\apj]
  {10.1088/0004-637X/808/2/126}, \href
  {https://ui.adsabs.harvard.edu/abs/2015ApJ...808..126V} {808, 126}

\bibitem[\protect\citeauthoryear{{Weiss} \& {Schlattl}}{{Weiss} \&
  {Schlattl}}{2008}]{2008Ap&SS.316...99W}
{Weiss} A.,  {Schlattl} H.,  2008, \mn@doi [\apss] {10.1007/s10509-007-9606-5},
  \href {https://ui.adsabs.harvard.edu/abs/2008Ap&SS.316...99W} {316, 99}

\bibitem[\protect\citeauthoryear{{Wheatley} et~al.,}{{Wheatley}
  et~al.}{2018}]{2018MNRAS.475.4476W}
{Wheatley} P.~J.,  et~al., 2018, \mn@doi [\mnras] {10.1093/mnras/stx2836},
  \href {https://ui.adsabs.harvard.edu/abs/2018MNRAS.475.4476W} {475, 4476}

\bibitem[\protect\citeauthoryear{{Zahn}}{{Zahn}}{1989}]{1989A&A...220..112Z}
{Zahn} J.~P.,  1989, \aap, \href
  {https://ui.adsabs.harvard.edu/abs/1989A&A...220..112Z} {220, 112}

\bibitem[\protect\citeauthoryear{{Zahn} \& {Bouchet}}{{Zahn} \&
  {Bouchet}}{1989}]{1989A&A...223..112Z}
{Zahn} J.~P.,  {Bouchet} L.,  1989, \aap, \href
  {https://ui.adsabs.harvard.edu/abs/1989A&A...223..112Z} {223, 112}

\makeatother
\end{thebibliography}

\bsp	
\label{lastpage}
\end{document}